\documentclass[3p, twocolumn]{elsarticle}

\usepackage{lineno,hyperref}
\modulolinenumbers[5]

\usepackage{siunitx}
\usepackage{graphicx}
\usepackage{verbatim}
\usepackage{color}
\usepackage{tabu}

\journal{Journal of \LaTeX\ Templates}

\begin{document}

\begin{frontmatter}

\title{Radiation Hardness Test of Eljen EJ-500 Optical Cement}

\author[lbladd,cal]{N. Buechel}

\author[lbladd]{S. Garrett}

\author[lbladd,mikeaff]{M. Lomnitz}

\author[lbladd]{A. Schmah\corref{mycorrespondingauthor}}
\cortext[mycorrespondingauthor]{Corresponding author}
\ead{aschmah@lbl.gov}

\author[lbladd,normal]{X. Sun}

\author[lbladd,cal]{J. Van Dyke}

\author[lbladd,normal]{J. Xu}

\author[lbladd,jinlongaff]{J. Zhang}

\address[lbladd]{Lawrence Berkeley National Lab, 1 Cyclotron Road, Berkeley, CA 94720}
\address[mikeaff]{Kent State University, 800 East Summit Street, Kent, OH 44240}
\address[jinlongaff]{Institute of Modern Physics Lanzhou, 509 Nanchang Road, Lanzhou, Gansu 730000, China}
\address[normal]{Central China Normal University, 152 Luoyo Rd, Hongshan, Wuhan, Hubei 430079, China}
\address[cal]{University of California: Berkeley, Berkeley, CA 94720}

\begin{abstract}
We present a comprehensive account of the proton radiation hardness of Eljen Technology's EJ-500 optical cement used in the construction of experiment detectors. The cement was embedded into five plastic scintillator tiles which were each exposed to one of five different levels of radiation by a 50 MeV proton beam produced at the 88-Inch Cyclotron at Lawrence Berkeley National Laboratory. A cosmic ray telescope setup was used to measure signal amplitudes before and after irradiation. Another post-radiation measurement was taken four months after the experiment to investigate whether the radiation damage to the cement recovers after a short amount of time. We verified that the radiation damage to the tiles increased with increasing dose but showed significant improvement after the four months time interval.
\end{abstract}


\end{frontmatter}


\section{Introduction}
Optical cement is a common element of scintillator based detector construction used for holding various components of such a detector in place while allowing for photons to pass through relatively uninhibited. The Event Plane and Centrality Detector (EPD), to be installed and used during RHIC Beam Energy Scan Phase II (BESII), utilizes Eljen Technology's EJ-500 optical cement in its construction to hold optical fibers in the constituent tiles. This improves the optical contact and allows for better signal collection. However, it opens up the question of how the optical cement, along with the optical fiber and the scintillator itself, respond to the radiation environment that is expected. We estimated that the amount of radiation during BESII in collider mode is on the order of $3\times10^{11}$ ions/cm$^{2}$. The additional background radiation is still unknown. It is necessary to ensure that radiation damage does not affect our measurements within the two years of BESII running, as well as for any additional runs that utilize the EPD. 

There is no documentation of previous radiation hardness experiments with EJ-500 cement. However, Bicron-600 made by Saint-Gobain S.A. is a similar epoxy that has been shown to sustain critical damage when exposed to a high amount of gamma radiation~\cite{kirn1999absorption}. These test results are not applicable to the EPD due to the fact that the measurement was conducted with a single dose of gamma radiation greatly exceeding our expected radiation limit. We found it necessary to conduct a dedicated experiment on EJ-500 to determine its durability under the conditions expected during BESII.


\section{Test Components}
The materials for the EPD construction consist of 1 cm thick Eljen EJ-200 plastic scintillator, split into trapezoidal tiles with Kuraray Y-11(200) polystyrene Wavelength Shifting (WLS) fibers embedded and glued into grooves in each tile with Eljen EJ-500 optical cement. For the radiation experiment we used five identical tiles assembled in this way in order to verify the effects of radiation in a practical setting.

\begin{figure}[h!]
\centering
\includegraphics[width=0.22\textwidth]{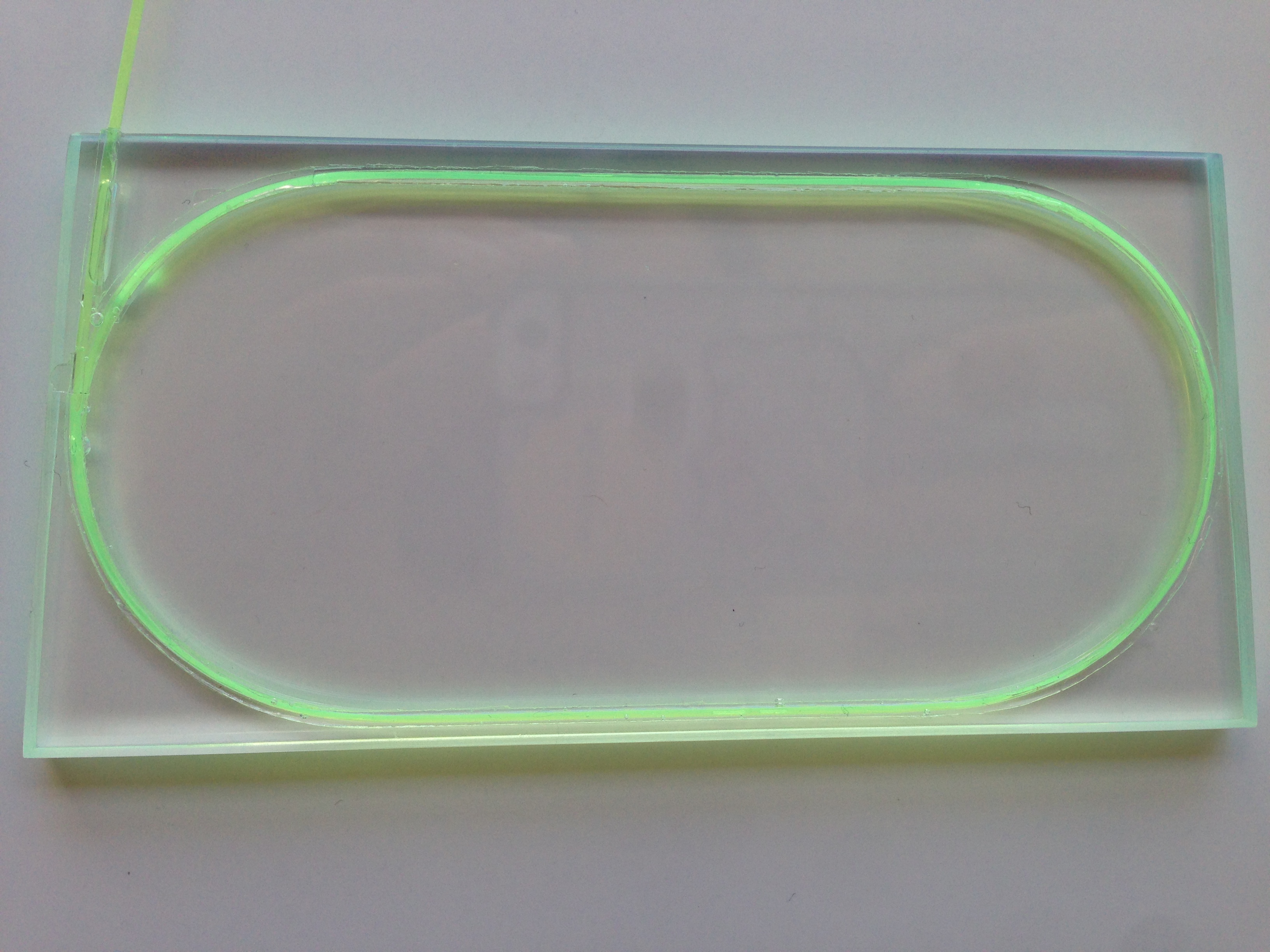}
\includegraphics[width=0.22\textwidth]{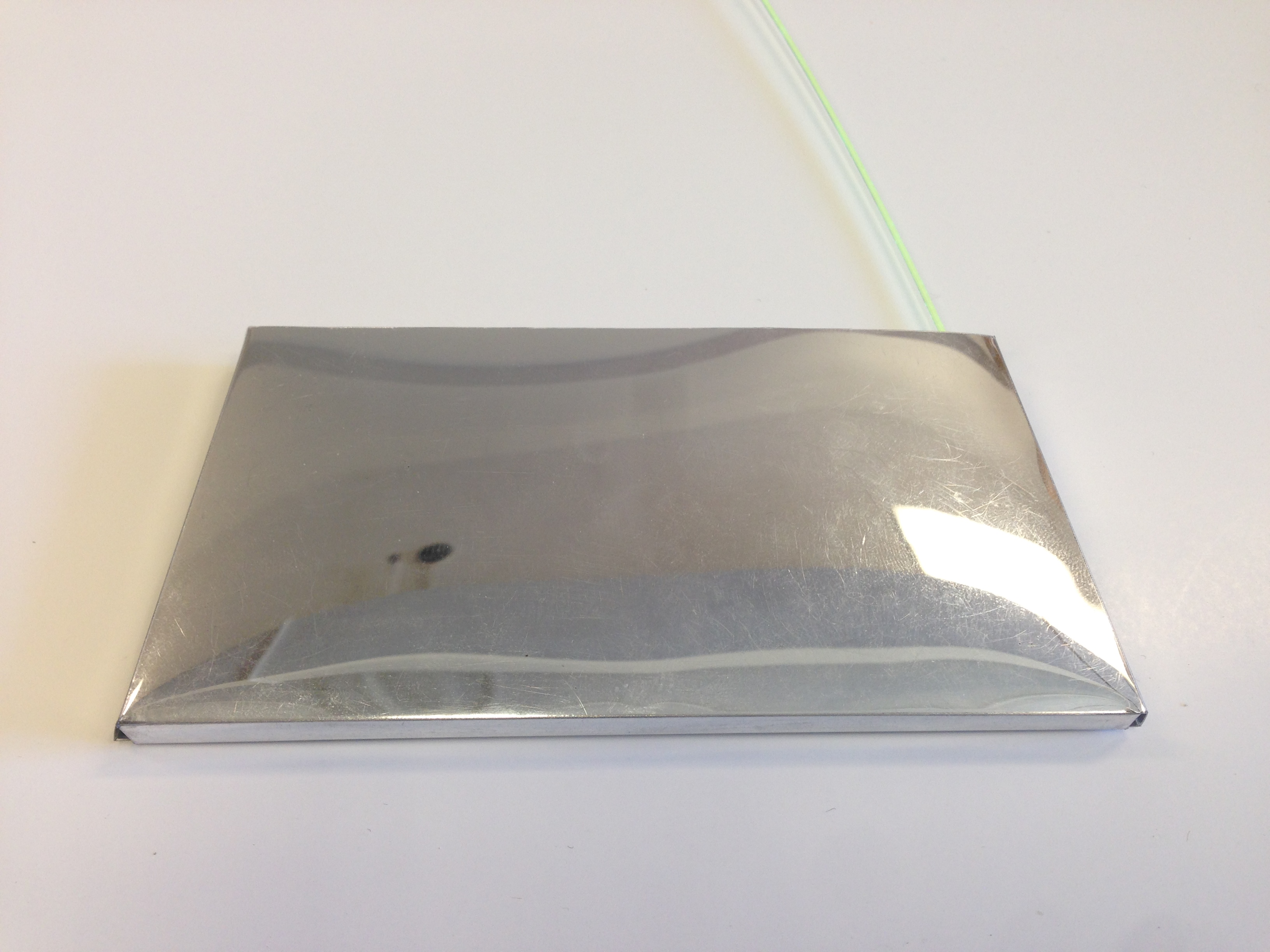}
\caption{12cm$\times$5.5cm$\times$0.5cm EJ-200 scintillator with a triple layer of Kuraray Y-11 WLS fibers glued with EJ-500 optical cement. This was one of the tiles used in the radiation hardness experiment. The right picture shows the same tile wrapped in aluminum mylar.}
\label{fig:tile}
\end{figure}

\paragraph{Scintillator Tiles}
The design of the test tiles was directly taken from that of a midrange tile in the EPD. The tiles were machined from EJ-200 plastic scintillator sheets into  $12$ cm$\times5.5$ cm$\times0.5$ cm trapezoids. Sigma-shaped grooves 0.16 cm wide were machined into the tile to a depth of 0.36 cm in order to hold three layers of a single WLS fiber for each tile (See below section on Optical Fibers for more information). An assembled test tile is shown in Figure \ref{fig:tile}. 

All machined edges of the tiles required polishing in order to minimize the corruption of the signal by reflection or absorption in the rough surfaces. We found that the most satisfactory polish came as a result of hand-polishing with a combination of fine grit sandpaper and aluminum oxide powder with a grit size on the order of \SI{5}{\micro\metre}. The fiber grooves were polished using wet sandpaper and a thin metal tool to fit in the groove.

Once the edges were sufficiently polished, each tile was cleaned and the optical fibers were glued into the sigma grooves (see below section on Epoxy Preparation). After the epoxy hardened the tiles were cleaned again before being wrapped in aluminum mylar to contain the signal photons inside the scintillator. We found that simply hand-wrapping a piece of mylar around the tiles was insufficient because the result was not tight enough. Instead, we drew a template cut for the mylar that would minimize the total surface area of the mylar while completely covering all faces of the tile. The mylar was then laid over the template and cut out with a box cutter for each tile. Any piece of the mylar that would be a fold on the tile was also perforated somewhat so as to ensure sharp edges all around. A wrapped test tile is shown in Figure \ref{fig:tile} (right). This mylar wrapping was used during the pre- and post-radiation measurements but was not in place during the actual irradiation of the tiles.

\paragraph{Optical Fibers}
The fibers used in the scintillator tiles for signal collection are Kuraray Y-11(200) Wavelength Shifting Fibers (WLS). These fibers have an attenuation length of $\lambda =400$ cm. In order to transmit as much light as possible the fibers need to have polished ends cut perpendicular to the fibers' longitudinal axes. A six-step process as outlined below was utilized to achieve the desired quality of the fiber ends.
\begin{enumerate}
\item The required lengths of fiber were cut from the spool. This step leaves the ends of the fibers jagged.

\item \label{itm:step2} A razor fiber cutting tool was used to cut the ends of the fibers. When done correctly, the fiber ends were smoothly cut perpendicular to the longitudinal axis of the fiber with minimal fraying of the encasing cladding.

\item \label{itm:step3} Polishing the ends of the fibers utilized four fiber polishing sheets with grits ranging from \SI{5}{\micro\metre} to \SI{0.3}{\micro\metre}. These sheets are sold by Thorlabs, with the \SI{5}{\micro\metre} grit size being the largest grit size. A plastic fiber holder was designed and fabricated with the 3D printer in order to maintain the fiber's perpendicular cut surface while polishing. The fibers were polished using a figure-eight pattern 30-40 times on each paper to achieve uniform quality.

\item \label{itm:step4} After polishing with the four lapping sheets the ends of the fibers had a smooth, mirror-like quality. In some cases one or two deep scratches were revealed. To remedy this, step \ref{itm:step3} was repeated until a satisfactory quality polish was achieved. Also, if too much pressure was applied to the fiber while polishing the cladding became frayed. In these cases it was often necessary to start over from step \ref{itm:step2}.
\end{enumerate}

After polishing, the ends of the WLS fibers that would be embedded in the scintillator were painted with with four layers of Eljen EJ-510 reflective paint in order to reflect back photons that head into the scintillator. Once the paint dried the WLS fibers were embedded and glued into the sigma grooves in the tiles. The test tiles were designed to hold a triple layer of WLS fiber in the sigma groove in order to produce a signal two times greater compared to a single layer of WLS fiber. A layer of optical epoxy was applied before and after each layer of WLS fiber in order to maximize the hold of the cement and the optical contact of the fiber with the scintillator.

\paragraph{Epoxy Preparation}
EJ-500 optical cement comes in two parts: a resin and a hardener, which are combined in a 4:1 ratio by volume to form the cement. In the process of mixing the two parts together, air pockets are formed in the epoxy which can degrade the signal quality of the detector. Before utilizing the EJ-500 optical cement to glue the fibers into place it was necessary to degas the epoxy to allow for better quality signal collection by the WLS fibers.

We estimated that we would need around 6 mL of epoxy to assemble the test tiles. After hand-mixing the resin and hardener together in a special container, we placed the sealed container in a special degasing centrifuge. The centrifuge first tumbled the container for 30 seconds to thoroughly mix the epoxy before spinning for 1.5 minutes to allow the air bubbles to rise to the surface. Afterwards, we opened the container and placed it in a vacuum pump chamber to draw out any leftover air pockets. The epoxy was left in vacuum until we could visibly discern that the amount of bubbles escaping the surface had drastically reduced and very few air pockets remained in the cement. The centrifuge and vacuum steps combined produced an optical cement virtually free from bubbles.

In order to keep the test tiles clean and minimize any waste of the optical cement we covered the top surface of the scintillator with Ultra-Slippery Tape Made with Teflon® PTFE. This tape adheres well to the surface of the scintillator, cuts easily and leaves no residue when removed. We formed a mask for the tiles from the tape by carefully cutting the tape out from the fiber grooves. We then applied a layer of the mixed epoxy inside the groove and put the first fiber layer in place. We repeated the step of applying the cement in between each layer of fiber and added an additional layer on top of the fiber to ensure that the space between the scintillator and fiber was entirely filled with optical cement. In total we used around 1.25 mL of EJ-500 epoxy per tile. We then removed the tape mask and applied more Ultra-Slippery Tape to certain parts of the fiber that threatened to spring out from the groove. The cement was then allowed to cure and harden for 24 hours.

\section{Test with Cosmic Rays}

\paragraph{Setup}
Once the five test tiles were fully assembled we measured the base signal of each tile with a cosmic ray telescope setup. This setup consisted of two $10$ cm$\times10$ cm$\times1$ cm tiles with embedded WLS fibers functioning as our trigger detectors with a test tile sandwiched in between. The two trigger detector tiles were assembled in a similar way to the test tiles, including the aluminum mylar wrapping, with the addition of black tape covering all surfaces to ensure that non-signal photons could not enter the scintillator. The entire setup excluding the power supply and oscilloscope was contained in a black box to enhance the light-tight aspect of the setup.

\paragraph{Electronics}
The signals of the tiles, both test and trigger, were read out by $1.3\times1.3$ mm Hamamatsu S13360-1325PE MPPCs (SiPMs) which were coupled to the wavelength shifting fibers extending from the tiles. The SiPMs were housed in custom 3D printed fiber-to-SiPM connectors to keep the connection secure and reduce signal noise. The SiPMs were then connected to Hamamatsu Photonics C12332 driver circuits for non-cooled MPPCs consisting of a sensor board and a power supply board~\cite{Hamamatsu}. The signals were read from the circuits by an oscilloscope.

\paragraph{Preliminary Data Collection}
In the two days before conducting the experiment we measured 400 individual trigger events per tile, about half of which displayed rapidly fluctuating signals. We excluded these ``bad" events from our data using an integral cut which left us with an average of 200 measurements per tile. For the measurements, we utilized a trigger threshold of 6 mV for the two reference detectors. These detectors exhibited typical signals on the order of 40 mV. The test tiles displayed average signals on the order of 50 mV. These results are in line with expected signals based on tile size as well as optical fiber geometry. 

\section{Test with Proton Beam}
\begin{figure}[h!]
\centering
\includegraphics[width=0.4\textwidth]{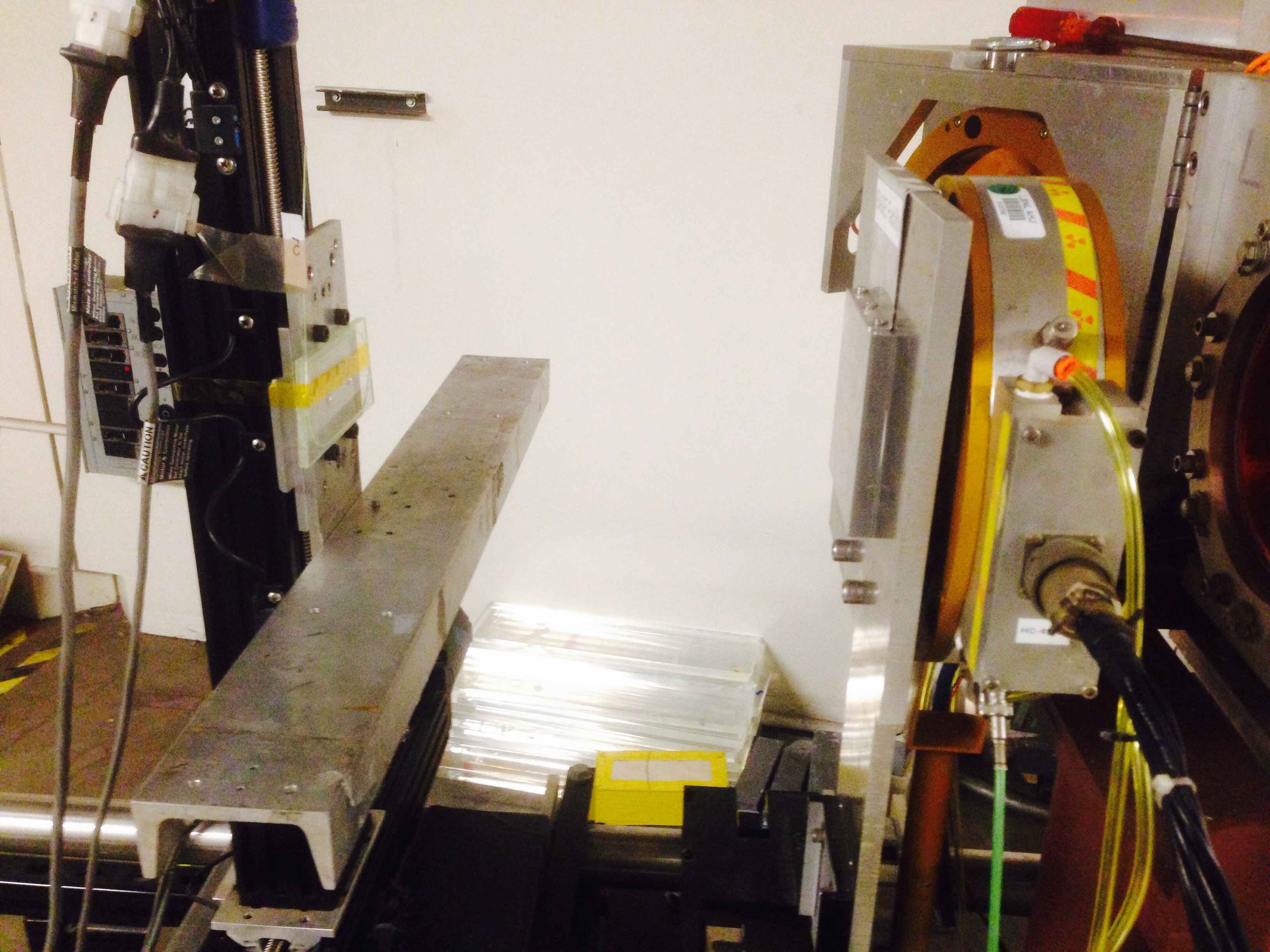}
\caption{Test tile attached to a vertical support post (left) in the direct path of the beamline, positioned about 25 cm away from the beam window (right).}
\label{fig:experiment}
\end{figure}

The radiation experiment was conducted at Lawrence Berkeley National Lab's 88-Inch Cyclotron with a 50 MeV proton beam. The tiles were positioned in a way that about 90\% of the surface was uniformly irradiated. A camera was set up inside the experiment vault so that we could monitor the tiles by live video feed as they were being irradiated. No bubbling or melting of any component was seen throughout the entire experiment. The tiles were attached one at a time to a vertical bar directly in the path of the beam. This setup is pictured in Figure \ref{fig:experiment}. Once the experiment finished all tiles were left in the experiment vault for two to four days to allow time for the activation levels in the tiles to decrease so that they could be safely handled and tested.

The tiles, starting with Tile \#5 and ending with Tile \#1, remained in the beam path until they reached the desired level of radiation. This took between 220 seconds to 4,100 seconds. The length of time in seconds as well as the corresponding radiation dosages for each tile are listed in Table \ref{table:1}. After irradiating Tile \#4 we removed the beam attenuators to allow for greater flux, shortening the amount of time each remaining tile spent in the beam. Tile \#2 received just above the expected amount of radiation for the two year duration of BESII. Tile \#1 was irradiated to three times the maximum amount of radiation expected.


\section{Analysis}
A new set of average and individual measurements were taken for each tile during the week after the experiment. Tiles 3, 4 and 5 were at a low enough activation level to be handled two days after the experiment. Tiles 1 and 2 reached safe handling levels four days after the experiment. The measurements were taken with the same setup and electronics as the measurements before the experiment for consistency. The same trigger threshold of 6 mV for the reference detectors was also used.

\paragraph{Signals}
The post-experiment measurements utilized the same trigger threshold of 6 mV for the reference detectors and the same integral cut on bad events. Once again we collected about 400 single events for each tile and were left with an average of 200 ``good" events per tile after the cut was applied. The average measurements of the test tiles showed that tiles 3, 4 and 5 still exhibited signals on the order of 50 mV. Tiles 1 and 2 dropped to average signals on the order of 45 mV. 

For our analysis of the change in individual event signal strengths, the signals per tile both before and after irradiation were corrected for event-by-event fluctuations due to lower or higher energy deposits by dividing the tile signal by the average of the trigger detector signals. We then calculated the light collection efficiency of each tile by dividing the tile's average signal after irradiation by the average signal before irradiation.

\section{Results}
With the exception of Tile \#5 which showed no change in efficiency, each of the other four test tiles showed a decrease in efficiency with increasing levels of radiation, with the sharpest decrease manifested in Tile \#1. Table \ref{table:1} shows the radiation levels for each test tile along with their calculated efficiency after radiation. We found that the efficiency of the tile falls to 70\% when irradiated with $1.0\times10^{12}$ ions/cm$^{2}$. As mentioned before, this level of radiation, corresponding to Tile \#1, is about 3 times the maximum amount of radiation expected during BES-II. We performed another event-by-event measurement and analysis on all tiles four months after the experiment to see if the efficiency would recover over time. We found that all tiles almost completely recovered from the radiation damage four months after irradiation. 

\begin{table*}[h!]
\centering
\fontsize{8}{5.2}\selectfont
\begin{tabu} to 0.8\textwidth { | X[l] || X[c] | X[c] | X[c] | X[c] | X[c] || }
 \hline
  & Tile 1 & Tile 2 & Tile 3 & Tile 4 & Tile 5\\
 \hline
 Time (s) & 4,097 & 1,962 & 431 & 768 & 216 \\
 Flux (ions/cm$^{2}$s) & $2.5\times10^8$ & $2.5\times10^8$ & $2\times10^8$ & $6.5\times10^7$ & $5\times10^7$ \\
 Fluence (ions/cm$^{2}$) & $1\times10^{12}$ & $5\times10^{11}$ & $1\times10^{11}$ & $5\times10^{10}$ & $1\times10^{10}$ \\
 Dose (kRad) & 300 & 150 & 30 & 15 & 3 \\
 Efficiency (\%) & 70 & 82 & 89 & 97 & 100\\
 \hline
\end{tabu}
\caption{Irradiation data per tile.}
\label{table:1}
\end{table*}

\section{Radiation Damage Attribution}
\begin{figure}[h!]
\centering
\resizebox{6cm}{!}{%
  \includegraphics[]{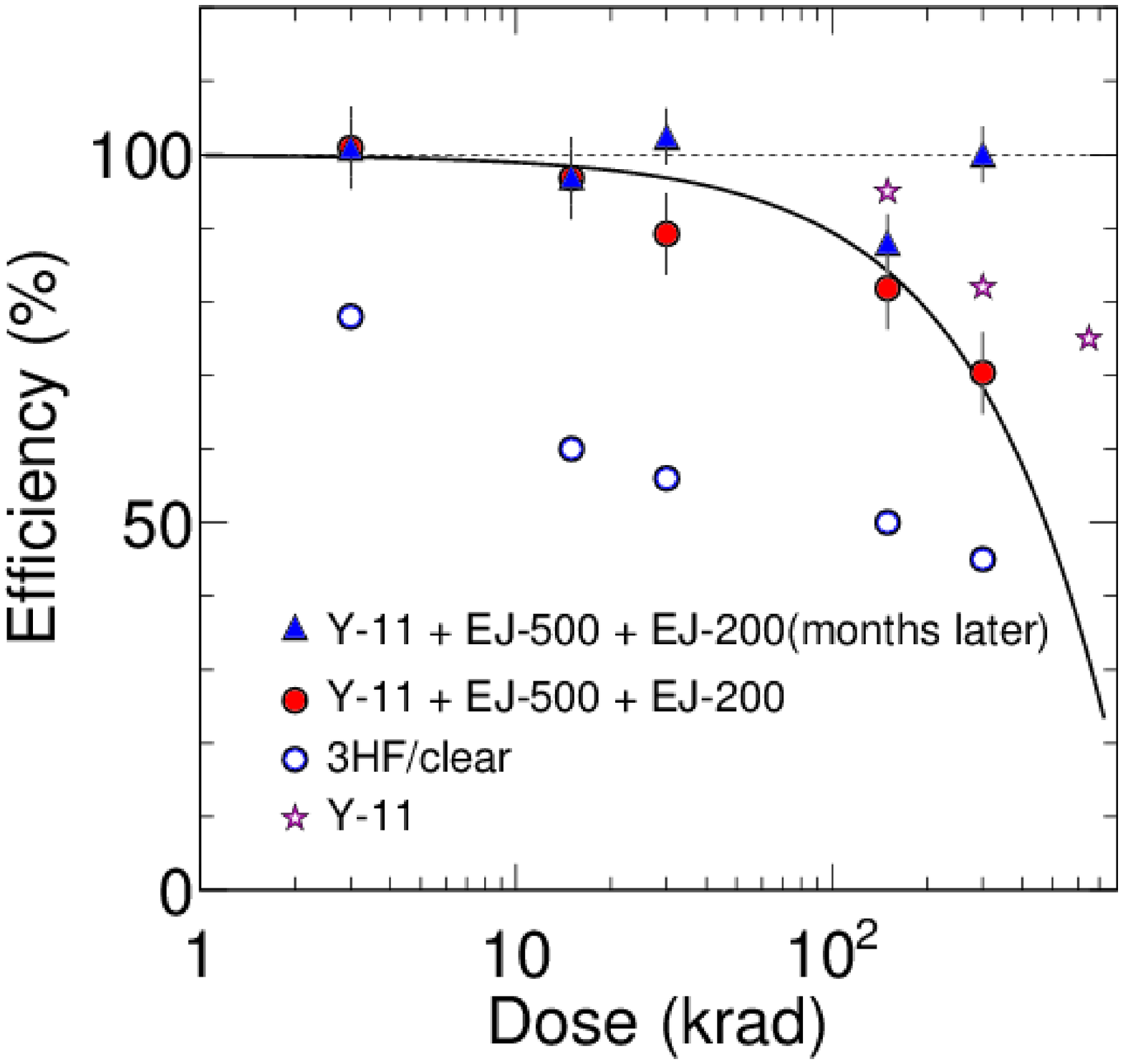}}
\caption{Measurement of the efficiency rate per dose of the assembled detector tiles compared with the reported efficiency rates of the Kuraray Y-11 WLS and 3HF clear optical fibers. EJ-500 is the optical cement while EJ-200 represents the plastic scintillator.}
\label{fig:results}
\end{figure}

To be conscientious of the fact that there are multiple components to our test tiles we researched radiation tests on the optical fibers that were used in their construction. There have been studies conducted on the radiation hardness of Kuraray WLS fibers that show agreement with our result of a 30\% loss in detector efficiency. One such study found that after a dose of 650 kRad the fibers had an efficiency of 70\%~\cite{Fleming:2012jf}. For our experiment, we had $1.0\times10^{12}$ protons/cm$^{2}$ in our maximum irradiated tile, while the average energy deposited in the fibers during the experiment is calculated to be 19 MeV $\cdot$ cm$^{2}$/g. This equates to a maximum dosage of 300 kRad in our WLS fibers, a factor of two less than the aforementioned study. A second study measured the attenuation length of Kuraray’s 3HF clear optical fibers to decrease by about 50\% after an irradiation dosage of 300 kRad~\cite{KHara:1998kh}. The results of both studies are compared to our experiment results in Figure \ref{fig:results}. Both studies report that while a small percentage of the radiation damage to the fibers is permanent, the fibers are able to partially recover from the damage after a rest time on the order of one to two months. From the comparison of our 30\% loss in efficiency with the aforementioned studies we conclude that our result is in agreement with the reduction in efficiency of WLS fibers that have been exposed to high dosages of radiation. This indicates that the WLS fibers account for the majority of the radiation damage to the tiles. With the efficiency back to normal after about four months we consider the whole detector setup to be safe for BESII.

\section{Summary}
We measured the radiation damage to Eljen EJ-500 optical cement in mock scintillator detectors using a 50 MeV proton beam at the 88-Inch Cyclotron at Lawrence Berkeley National Lab. We found that the efficiency of a detector tile stays constant at a radiation fluence of $1.0\times10^{10}$ protons/cm$^{2}$, but decreases by 30\% when irradiated at $1.0\times10^{12}$ protons/cm$^{2}$. This efficiency reduction recovers after about four months of rest time. Our result agrees with the reduction in efficiency seen in Kuraray WLS optical fibers after they are inundated with high dosages of radiation. This is indicative that a majority of the radiation damage to the tiles can be attributed to the WLS fibers. Consequently, we have determined that the optical epoxy suffers no noticeable impairment from the amount of radiation applied during the experiment. In conclusion, Eljen EJ-500 optical cement can be considered radiation hard for radiation levels at or below $1.0\times10^{12}$ protons/cm$^{2}$.

\section{Acknowledgments}
We would like to thank Kenneth Wilson and Thomas Johnson from the LBL Engineering Department for their assistance with degassing the optical cement. We thank the LBL 88-Inch Cyclotron staff for their support during the radiation experiment.

\bibliography{mybibfile.bib}
\bibliographystyle{unsrt}

\end{document}